# Analogue gravity in hyperbolic metamaterials

Igor I. Smolyaninov

*Department of Electrical and Computer Engineering, University of Maryland, College Park, MD 20742, USA*

**Sub-wavelength confinement of light in nonlinear hyperbolic metamaterials due to formation of spatial solitons has attracted much recent attention because of its seemingly counter-intuitive behavior. In order to achieve self-focusing in a hyperbolic wire medium, a nonlinear self-defocusing Kerr medium must be used as a dielectric host. Here we demonstrate that this behavior finds natural explanation in terms of analogue gravity. Wave equation describing propagation of extraordinary light inside hyperbolic metamaterials exhibits 2+1 dimensional Lorentz symmetry. The role of time in the corresponding effective 3D Minkowski spacetime is played by the spatial coordinate aligned with the optical axis of the metamaterial. Nonlinear optical Kerr effect "bends" this spacetime resulting in effective gravitational force between extraordinary photons. In order for the effective gravitational constant to be positive, negative self-defocusing Kerr medium must be used as a host. If gravitational self-interaction is strong enough, spatial soliton may collapse into a black hole analogue.**

Light propagation through hyperbolic metamaterials has attracted much recent attention due to their ability to guide and manipulate electromagnetic fields on a spatial scale much smaller than the free space wavelength [1-6]. Almost immediately it was realized that nonlinear optical effects may further increase light confinement in hyperbolic metamaterials based on metal nanowires and metal nanolayers immersed in a Kerr-type dielectric host (Fig.1). Several reports demonstrated existence of spatial solitons in an array of nanowires embedded in a Kerr medium [7-10]. An interesting counter-intuitive feature of these solitons is that they occur only if a self-defocusing Kerr medium is used as a dielectric host. These interesting results have been obtained based on either coupled-mode theory [7,8], or on a detailed analysis of trapped states in nonlinear hyperbolic media [9]. While definitely valid, these approaches provide limited physical insight of this and other more general problems of soliton interaction in nonlinear hyperbolic metamaterials. In this paper we demonstrate that behavior of spatial solitons in nonlinear hyperbolic metamaterials finds natural explanation in terms of analogue gravity. Wave equation describing propagation of extraordinary light inside hyperbolic metamaterials exhibits 2+1 dimensional Lorentz symmetry. The role of time in the corresponding effective 3D Minkowski spacetime is played by the spatial coordinate aligned with the optical axis of the metamaterial. Nonlinear optical Kerr effect "bends" this spacetime resulting in effective gravitational force between extraordinary photons. In order for the effective gravitational constant to be positive, negative self-defocusing Kerr medium must be used as a host. While this observation explains basic findings of refs.[7-10], it has much more general consequences for soliton physics in hyperbolic metamaterials, since both classic and quantum gravity in 2+1 spatial dimensions is known to be an exactly soluble system [11].

As a starting point, let us recall basic features of hyperbolic metamaterial modeling using 2+1 dimensional Minkowski spacetime. Recent advances in electromagnetic metamaterials enable design of novel physical systems which can be described by effective space-times having very unusual metric and topological properties. In particular, hyperbolic metamaterials offer an interesting experimental window into physics of Minkowski spacetimes, since propagation of extraordinary light inside a hyperbolic metamaterial is described by wave equation exhibiting 2+1 dimensional Lorentz symmetry [12]. The role of time in the corresponding effective 3D Minkowski spacetime is played by the spatial coordinate, which is oriented along the optical axis of the metamaterial [13]. This spacetime may be made "causal" by breaking the mirror and temporal symmetries of the metamaterial, which results in one-way light propagation along the "timelike" spatial coordinate [14]. Two different Minkowski spacetimes may "collide" forming a Minkowski domain wall [15,16], while "bending" the effective spacetime may lead to an experimental model of the big bang [13]. Under certain circumstances, thermal fluctuations of the metamaterial may cause transient formation of hyperbolic regions (3D Minkowski spacetimes) inside the metamaterial [17], so that the resulting picture of multiple transient Minkowski spacetimes looks somewhat similar to cosmological multiverse. Interestingly, the physical vacuum itself behaves as a hyperbolic metamaterial when subjected to very strong magnetic field [18,19]. Therefore, the Minkowski spacetime analogues mentioned above appear to be quite meaningful. Despite this rich and interesting physics, all the metamaterial-based Minkowski spacetime models described so far were limited in one very important respect: the effective metric was decoupled from the matter content (photons) of these spacetimes. Here we consider nonlinear optical effects which "bend" the effective Minkowski spacetime resulting in gravity-like interaction of extraordinary photons. We demonstrate that nonlinear optical Kerr effect results in effective gravitational force between extraordinary photons. If gravitational self-interaction is strong enough, spatial soliton may collapse into a black hole analogue.

First, let us demonstrate that the wave equation describing propagation of monochromatic extraordinary light inside a hyperbolic metamaterial does indeed exhibit 2+1 dimensional Lorentz symmetry. A detailed derivation of this result can be found in refs.[12,13]. We assume that the metamaterial in question is uniaxial and non-magnetic ($\mu$=1), so that electromagnetic field inside the metamaterial may be separated into ordinary and extraordinary waves (vector $\vec{E}$ of the extraordinary light wave is parallel to the plane defined by the *k*–vector of the wave and the optical axis of the metamaterial). Since hyperbolic metamaterials exhibit strong temporal dispersion, we will work in the frequency domain and assume that in some frequency band around $\omega=\omega_0$ the metamaterial may be described by anisotropic dielectric tensor having the diagonal components $\varepsilon_{xx}=\varepsilon_{yy}=\varepsilon_1 >0$ and $\varepsilon_{zz}=\varepsilon_2<0$. In the linear optics approximation all the non-diagonal components are assumed to be zero. Propagation of extraordinary light in such a metamaterial may be described by a coordinate-dependent wave function $\varphi_\omega=E_z$ obeying the following wave equation [12,13]:

$$-\frac{\omega^2}{c^2}\varphi_\omega = \frac{\partial^2\varphi_\omega}{\varepsilon_1\partial z^2} + \frac{1}{\varepsilon_2}\left(\frac{\partial^2\varphi_\omega}{\partial x^2}+\frac{\partial^2\varphi_\omega}{\partial y^2}\right) \qquad (1)$$

This wave equation coincides with the Klein-Gordon equation for a massive scalar field $\varphi_\omega$ in 3D Minkowski spacetime:

$$-\frac{\partial^2\varphi_\omega}{\varepsilon_1\partial z^2} + \frac{1}{(-\varepsilon_2)}\left(\frac{\partial^2\varphi_\omega}{\partial x^2}+\frac{\partial^2\varphi_\omega}{\partial y^2}\right)=\frac{\omega_0^2}{c^2}\varphi_\omega = \frac{m^{*2}c^2}{\hbar^2}\varphi_\omega \qquad (2)$$

in which spatial coordinate $z=\tau$ behaves as a "timelike" variable. Eq.(2) describes world lines of massive particles which propagate in a flat 2+1 dimensional Minkowski spacetime [12,13]. Note that components of metamaterial dielectric tensor define the effective metric $g_{ik}$ of this spacetime: $g_{00}=-\varepsilon_1$ and $g_{11}=g_{22}=-\varepsilon_2$.

When the nonlinear optical effects become important, they are described in terms of various order nonlinear susceptibilities $\chi^{(n)}$ of the metamaterial:

$$D_i = \chi^{(1)}_{ij}E_j + \chi^{(2)}_{ijl}E_jE_l + \chi^{(3)}_{ijlm}E_jE_lE_m + ... \qquad (3)$$

Taking into account these nonlinear terms, the dielectric tensor of the metamaterial (which defines its effective metric) may be written as

$$\varepsilon_{ij} = \chi_{ij}^{(1)} + \chi_{ijl}^{(2)} E_l + \chi_{ijlm}^{(3)} E_l E_m + ... \quad (4)$$

It is clear that eq.(4) provides coupling between the matter content (photons) and the effective metric of the metamaterial "spacetime". However, in order to emulate gravity, the nonlinear susceptibilities $\chi^{(n)}$ of the metamaterial need to be engineered in some particular way. This task may be quite complicated due to effects of spatial dispersion, which may become prominent in hyperbolic metamaterials [19]. High frequency macroscopic electrodynamics of metamaterials may be described using two equivalent languages [20]. We can either introduce magneto-electric moduli relating (D,H) and (E,B) pairs:

$$\vec{D} = \vec{\varepsilon}\vec{E} + \vec{\alpha}\vec{B}, \quad (5)$$

$$\vec{H} = \vec{\mu}^{-1}\vec{B} + \vec{\beta}\vec{E},$$

or assume that $\vec{D} = \vec{\varepsilon}\vec{E}$ and $\vec{H} = \vec{B}$, while tensor $\vec{\varepsilon}(\omega_0, \vec{k})$ exhibits linear (odd) terms in spatial dispersion:

$$\varepsilon_{ij}(\omega_0, \vec{k}) = \varepsilon_{ij}(\omega_0, 0) + \gamma_{ijl}^{(1)} k_l + \gamma_{ijlm}^{(2)} k_l k_m + ..., \quad (6)$$

Connection between these two descriptions is easy to establish for time harmonic plane waves, since

$$\vec{B} = \frac{c}{i\omega} rot\vec{E} = \frac{c}{\omega}\left[\vec{k} \times \vec{E}\right], \quad (7)$$

When nonlinear optical effects need to be taken into account, the second choice is more convenient. Thus, we are going to assume that in the most general case all the $\chi^{(n)}$ terms in eq.(4) may depend on the photon wave vectors. This general framework encompasses all kinds of effective gravity theories, which are much more complicated than usual general relativity. Our goal is to find what kind of simplifications of this general framework may lead to metamaterial models which emulate usual gravity.

In the weak gravitational field limit the Einstein equation

$$R_i^k = \frac{8\pi\gamma}{c^4}\left(T_i^k - \frac{1}{2}\delta_i^k T\right) \tag{8}$$

is reduced to

$$R_{00} = \frac{1}{c^2}\Delta\phi = \frac{1}{2}\Delta g_{00} = \frac{8\pi\gamma}{c^4}T_{00} \,, \tag{9}$$

where $\phi$ is the gravitational potential [21]. Since in our effective Minkowski spacetime $g_{00}$ is identified with $-\varepsilon_1$, comparison of eqs. (4) and (9) indicates that all the second order nonlinear susceptibilities $\chi^{(2)}_{ijl}$ of the metamaterial must be equal to zero, while the third order terms may provide correct coupling between the effective metric and the energy-momentum tensor. These terms are associated with the optical Kerr effect. All the higher order $\chi^{(n)}$ terms must be zero at $n>3$.

Indeed, detailed analysis indicates that Kerr effect in a hyperbolic metamaterial leads to effective gravity. Since $z$ coordinate plays the role of time, while $g_{00}$ is identified with $-\varepsilon_1$, eq.(9) must be translated as

$$-\Delta^{(2)}\varepsilon_1 = \frac{16\pi\gamma^*}{c^4}T_{zz} = \frac{16\pi\gamma^*}{c^4}\sigma_{zz} \,, \tag{10}$$

where $\Delta^{(2)}$ is the 2D Laplacian operating in the $xy$ plane, $\gamma^*$ is the effective "gravitational constant", and $\sigma_{zz}$ is the $zz$ component of the Maxwell stress tensor of the electromagnetic field in the medium:

$$\sigma_{zz} = \frac{1}{4\pi}\left(D_z E_z + H_z B_z - \frac{1}{2}(\vec{D}\vec{E} + \vec{H}\vec{B})\right) \tag{11}$$

Let us find a contribution to $\sigma_{zz}$, which is made by a single extraordinary plane wave propagating inside the hyperbolic metamaterial. Assuming without a loss of generality that the *B* field of the wave is oriented along *y* direction, the other field components may be found from Maxwell equations as

$$k_z B_y = \frac{\omega}{c}\varepsilon_1 E_x, \quad \text{and} \quad k_x B_y = -\frac{\omega}{c}\varepsilon_2 E_z \qquad (12)$$

Taking into account the dispersion law of the extraordinary wave

$$\frac{\omega^2}{c^2} = \frac{k_z^2}{\varepsilon_1} + \frac{k_x^2 + k_y^2}{\varepsilon_2}, \qquad (13)$$

the contribution to $\sigma_{zz}$ from a single plane wave is

$$\sigma_{zz} = -\frac{c^2 B^2 k_z^2}{4\pi\omega^2 \varepsilon_1} \qquad (14)$$

Thus, for a single plane wave eq.(10) may be rewritten as

$$-\Delta^{(2)}\varepsilon_1 = -\Delta^{(2)}\left(\varepsilon_1^{(0)} + \delta\varepsilon_1\right) = k_x^2 \delta\varepsilon_1 = -\frac{4\gamma * B^2 k_z^2}{c^2 \omega^2 \varepsilon_1}, \qquad (15)$$

where we have assumed that nonlinear corrections to $\varepsilon_1$ are small, so that we can separate $\varepsilon_1$ into the constant background value $\varepsilon_1^{(0)}$ and weak nonlinear corrections. These nonlinear corrections do indeed look like the Kerr effect assuming that the extraordinary photon wave vector components are large compared to $\omega/c$:

$$\delta\varepsilon_1 = -\frac{4\gamma * B^2 k_z^2}{c^2 \omega^2 \varepsilon_1 k_x^2} \approx \frac{4\gamma * B^2}{c^2 \omega^2 \varepsilon_2} = \chi^{(3)} B^2 \qquad (16)$$

The latter assumption has to be the case indeed if extraordinary photons may be considered as classic "particles". Eq.(16) establishes connection between the effective gravitational constant $\gamma^*$ and the third order nonlinear susceptibility $\chi^{(3)}$ of the hyperbolic metamaterial. Since $\varepsilon_{xx}= \varepsilon_{yy}= \varepsilon_1 >0$ and $\varepsilon_{zz} = \varepsilon_2<0$, the sign of $\chi^{(3)}$ must be negative for the effective gravity to be attractive. For a metal wire array metamaterial shown in Fig.1(b) the diagonal components of the dielectric tensor may be obtained using Maxwell-Garnett approximation [22]:

$$\varepsilon_2 = \varepsilon_z = n\varepsilon_m + (1-n)\varepsilon_d \qquad (17)$$

$$\varepsilon_1 = \varepsilon_{x,y} = \frac{2n\varepsilon_m\varepsilon_d + (1-n)\varepsilon_d(\varepsilon_d + \varepsilon_m)}{(1-n)(\varepsilon_d + \varepsilon_m) + 2n\varepsilon_d} \qquad (18)$$

where $n$ is the volume fraction of the metallic phase (assumed to be small), and $\varepsilon_m$ and $\varepsilon_d$ are the dielectric permittivities of the metal and dielectric phase, respectively. Since $-\varepsilon_m \gg \varepsilon_d$, eq.(18) may be simplified as

$$\varepsilon_1 \approx \varepsilon_d \frac{(1+n)}{(1-n)} \sim \varepsilon_d \qquad (19)$$

Thus, we have recovered the main result of refs.[7-9]: in order to obtain attractive effective gravity the dielectric host medium must exhibit negative (self-defocusing) Kerr effect. Extraordinary light rays in such a medium will behave as 2+1 dimensional world lines of self-gravitating bodies and may collapse into sub-wavelength spatial solitons.

Let us modify eq.(2) by taking into account self-defocusing Kerr effect of the dielectric host. Assuming a spatial soliton-like solution which conserves energy per unit length $W \sim P/c$ (where $P$ is the laser power), the soliton width $\rho$ and the magnetic field amplitude $B$ are related as

$$B^2\rho^2 = P/c \qquad (20)$$

As a result, eq.(2) must be re-written as

$$-\frac{\partial^2 \varphi_\omega}{\left(\varepsilon_1^{(0)} - \frac{(-\chi^{(3)})P}{c\rho^2}\right)\partial z^2} + \frac{1}{(-\varepsilon_2)}\left(\frac{\partial^2 \varphi_\omega}{\partial x^2} + \frac{\partial^2 \varphi_\omega}{\partial y^2}\right) = \frac{\omega_0^2}{c^2}\varphi_\omega \qquad (21)$$

where $\varepsilon_1^{(0)}$ is the dielectric permittivity component at $P=0$ (note that nonlinear contribution to $\varepsilon_2 \approx n\varepsilon_m$ may be neglected). Effective metric described by eq.(21) has a black hole-like singularity at

$$\rho = \left(\frac{(-\chi^{(3)})P}{c\varepsilon_1^{(0)}}\right)^{1/2} \qquad (22)$$

Let us evaluate if this singularity may be observed in experiment. In order to be observable, the critical value described by eq.(22) must be larger than the metamaterial structure parameter (the inter-wire distance). On the other hand, light intensity must be small enough, so that higher order nonlinear effects may be neglected. It is obvious that negative Kerr effect in natural dielectrics is not strong enough to observe this singularity. On the other hand, artificial self-defocusing dielectrics having low $\varepsilon_d^{(0)}$ may be engineered based on nanoparticle suspensions in liquids. Such suspensions are widely available commercially. Because of the large and negative thermo-optic coefficient inherent to most liquids, heating produced by partial absorption of the propagating beam translates into a significant decrease of the refractive index at higher light intensity. For example, reported thermo-optics coefficient of water reaches $\Delta n/\Delta T = -5.7 \times 10^{-4} K^{-1}$ [23]. Moreover, introducing absorbent dye into the liquid allows for increased thermal nonlinear response [24].

The dielectric properties of nanoparticles needed to obtain an artificial low $\varepsilon_d^{(0)}$ medium may be calculated using Maxwell-Garnett approximation [25]:

$$\frac{\varepsilon_d - \varepsilon_l}{\varepsilon_d + 2\varepsilon_l} = \alpha_n \frac{\varepsilon_n - \varepsilon_l}{\varepsilon_n + 2\varepsilon_l} \qquad (23)$$

where $\alpha_n$ is the volume fraction of the nanoparticles (assumed to be small), and $\varepsilon_n$ and $\varepsilon_l$ are the dielectric permittivities of nanoparticles and liquid, respectively. A low $\varepsilon_d^{(0)}$ medium is obtained if $\varepsilon_n \approx -2\varepsilon_l$. Therefore, either plasmonic oxides or nitrades would be the best nanoparticle choice in the visible and near infrared ranges [26]. Based on the thermo-optics coefficient of water, $\Delta\varepsilon_d$~0.1 may be obtained at quite realistic $\Delta T$~50K. Therefore, $\varepsilon_l^{(0)}$~0.1 may be used in eq.(22) to estimate the critical value of soliton radius as a function of laser power. This estimate is presented in Fig.2. The critical soliton radius at $P$=100W equals ~20 nm, which does not look completely unrealistic from the fabrication standpoint. While achieving such critical values with CW lasers seems implausible, using pulsed laser definitely looks like a realistic option since thermal damage produced by a pulsed laser typically depends on the pulse energy and not the pulse power.

In conclusion, we have considered nonlinear optical effects in hyperbolic metamaterials, and demonstrated that negative nonlinear optical Kerr effect results in effective gravitational force between extraordinary photons. Our observation explains formation of subwavelength spatial solitons in metal wire array hyperbolic metamaterials based on self-defocusing dielectric host. The proposed physical mechanism has much more general consequences for soliton interaction in hyperbolic metamaterials, since both classic and quantum gravity in 2+1 spatial dimensions is known to be an exactly soluble system [11].

**Figure Captions**

**Figure 1.** (Color online) Typical geometries of nonlinear hyperbolic metamaterials: (a) multilayer metal-dielectric structure (b) metal wire array structure. The dielectric host $\varepsilon_d$ exhibits nonlinear optical Kerr effect.

**Figure 2**. Critical radius of spatial soliton calculated as a function of laser power for a metal wire array hyperbolic metamaterial built with an artificial self-defocusing liquid dielectric having $\varepsilon_l^{(0)} \sim 0.1$.

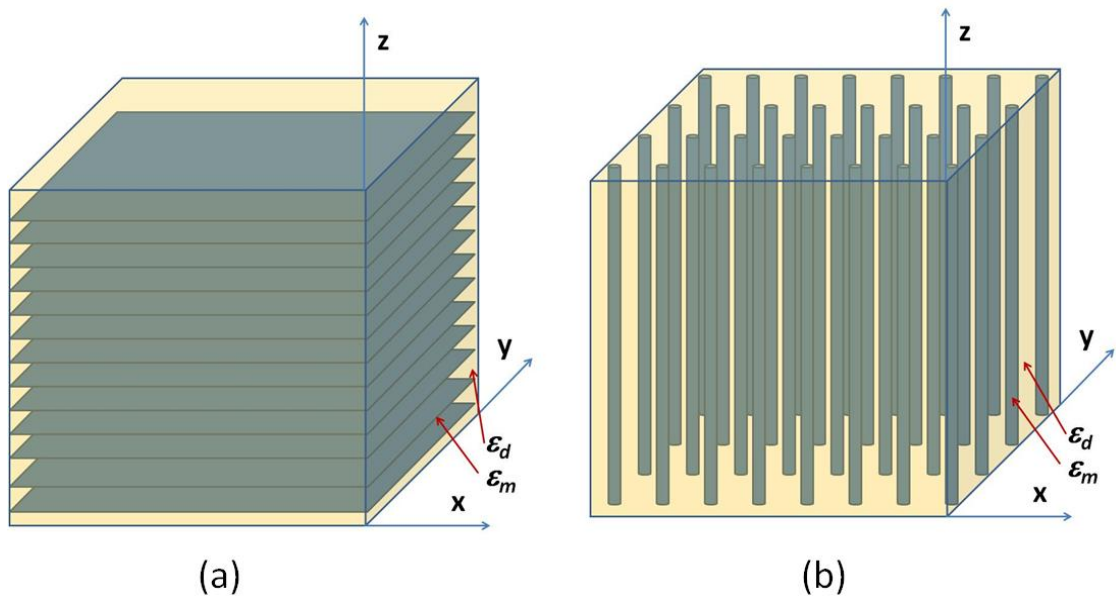

Fig. 1

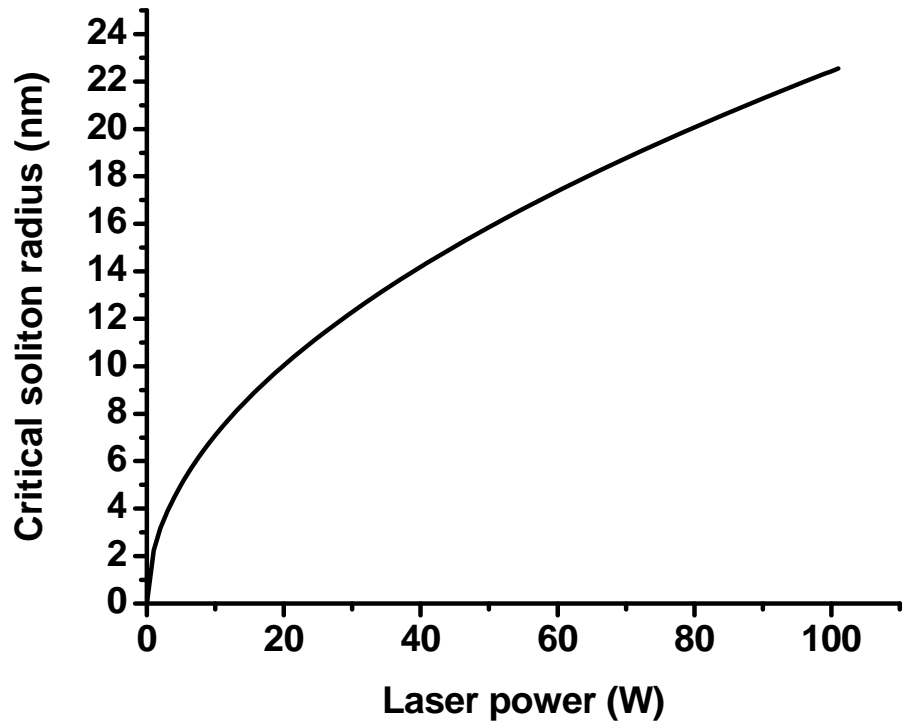

Fig. 2